\documentclass[doublecol]{epl2} 

\makeatletter
\renewcommand{\dash}{\texorpdfstring{\gdef\epl@dash{--}}{}}
\renewcommand{\godot}{\texorpdfstring{\gdef\epl@dot{.}}{}}
\makeatother

\usepackage{amssymb}
\usepackage{slashed}
\usepackage{graphicx}
\usepackage{graphics}
\usepackage{epsfig}
\usepackage{color}
\usepackage{soul}
\usepackage{tabularx}
\usepackage{amsmath}
\usepackage{amsfonts}
\usepackage{amssymb}
\usepackage{enumitem}
\usepackage{hyperref}
\usepackage{microtype}
\usepackage{verbatim}
\usepackage[normalem]{ulem}
\usepackage[utf8]{inputenc}

\usepackage{MnSymbol,wasysym}

\title{The Absolute Swampland}
\shorttitle{Absolute Swampland} 

\author{Astrid Eichhorn\inst{1} \and Arthur Hebecker\inst{2} \and Jan M. Pawlowski\inst{2} \and Johannes Walcher\inst{2,3}}
\shortauthor{Eichhorn, Hebecker, Pawlowski, Walcher}

\institute{                    
  \inst{1} CP3-Origins, University of Southern Denmark, Campusvej 55, 5230 Odense M, Denmark\\
  \inst{2} Institute for Theoretical Physics, Heidelberg University, Philosophenweg 16 $\&\!$ 19, 69120 Heidelberg, Germany\\
  \inst{3} Institute for Mathematics, Heidelberg University, Im Neuenheimer Feld 205, 69120 Heidelberg, Germany
}

\abstract{The ``Swampland Program'' aims to discriminate consistent-looking effective field theories that
do not admit a UV completion in quantum gravity from those that do. While most often developed under the 
umbrella of string theory, several swampland criteria have been explored also in other contexts, especially 
asymptotically safe gravity. A comparison between different approaches can help to clarify the dependence of 
low-energy constraints on UV physics and thereby shed light on the universality of quantum gravity 
itself. In this short review we summarise what is known about three important swampland conjectures in string theory 
and in asymptotic safety. We point out future lines of research that can help to understand to what extent swampland conjectures are absolute, i.e.~hold in quantum gravity in general, or relative, i.e.~belong only to a specific UV framework.}

\hypersetup{hidelinks}

\begin{document}

\maketitle

\section{Introduction and orientation}
\label{sec:Introduction}

The set of effective field theories (EFTs) that look consistent according to all available low-energy criteria, 
but do not arise from a UV complete theory that includes quantum gravity has become known as the swampland
\cite{Vafa:2005ui, Ooguri:2006in}. The basic idea is that despite the complexity of the set of vacua of string
theory, the swampland is much vaster still, and the low-energy physics observed in our Universe much less surprising
in comparison. Delineating the boundaries of the swampland has become an important part of research in 
quantum gravity.

A vast literature formulates, explores and attempts to prove swampland conjectures, see \cite{Brennan:2017rbf,Palti:2019pca,vanBeest:2021lhn,Grana:2021zvf,Agmon:2022thq} for reviews and references. While most of this work
is informed by string theory, it is of critical importance to understand whether the extracted criteria actually reflect
what is possible in quantum gravity or depend on the assumptions of this specific realisation. First discussions
of constraints across the boundaries of different UV completions and first results in asymptotically safe gravity appear, 
e.g., in \cite{deAlwis:2019aud, Basile:2021krr,Platania2024}.

\begin{figure*}
\begin{center}
\includegraphics[width=0.9\linewidth, clip=true, trim=0cm 14cm 0cm 0cm]{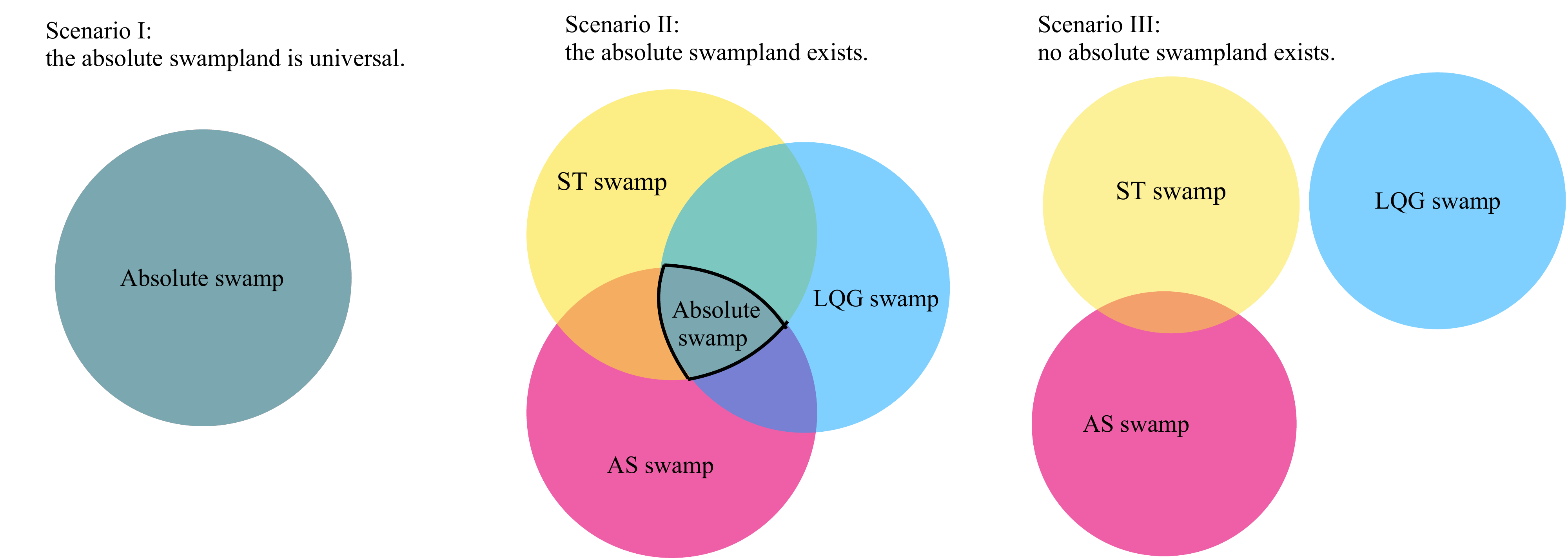}
\end{center}
\caption{\label{fig:sketch}  We sketch three distinct scenarios for the relation of relative swamplands to each other, using string theory (ST), asymptotic safety (AS) and Loop Quantum Gravity (LQG) as three examples for quantum-gravity theories. The absolute swampland is universal in scenario I.
No absolute swampland exists in scenario III, because the intersection of the stringy and asymptotically unsafe swampland is in the landscape of Loop Quantum Gravity. 
In scenarios I and II, absolute constraints on model-building exist, because effective field theories in the absolute swamp are not viable. In scenario II and III, observational tests of quantum gravity theories at low energy exist and consist in determining which relative swampland an EFT describing nature lies in.}
\end{figure*}

It is often assumed, implicitly or explicitly, that the swampland is universal, i.e., that all swampland criteria hold in all consistent approaches to quantum gravity, cf.~scenario I in Fig.~\ref{fig:sketch}. 
From the bottom-up perspective, it might indeed seem most satisfying to develop swampland criteria that hold completely 
independently of any particular assumptions about the microscopic degrees of freedom.
We aim to critically review this notion of a universal swampland.
We should then begin by asking, in each particular approach, 
for the set of EFTs that are not embeddable in that framework  and thus not compatible with the swampland criteria of this approach. We will refer to this as a \emph{relative} swampland.
The intersection of all relative swamplands forms the \emph{absolute} swampland. It consists of all EFTs that cannot be UV completed in any consistent approach to quantum gravity. From the pragmatic point of view of low-energy model-building, we should ultimately not care whether any given EFT can be UV completed in a particular approach, as long as this is possible in \emph{some} (at least one) of them, i.e., as long as the EFT does not lie in the absolute swampland.

We call the swampland universal iff all swampland criteria are shared between all quantum gravity theories, i.e., if the relative swamplands are identical with the absolute swampland.\footnote{The set 
of EFTs satisfying all  
swampland criteria  in all approaches is called the universal landscape, as they make sense 
independently of the framework for UV completion. This universal landscape is the complement of the union of all swamplands.} 

We can envision several scenarios\footnote{We understand that a similar set of options has recently been considered 
by A.\ Platania.} in the set of all EFTs, illustrated in Fig.~\ref{fig:sketch}. The 
situation 
 of a universal swampland (scenario I) is often implicitly assumed in the literature and 
would indicate a thorough understanding of the low-energy constraints
imposed by quantum gravity. It would in turn naturally beg the question whether there is in fact a deeper relation
between the underlying microscopic theories (similar to string dualities). The boundaries of such a universal swamp 
would provide sharp constraints on low-energy model building. 
On the other extreme it could be the case that 
the absolute swampland is \emph{empty}, in other words that every EFT admits at least one UV completion that 
includes gravity (scenario III). We personally find this unlikely, and it would be disappointing, as it would indicate that 
gravity really is not that special or constraining after all. The most likely, and exciting, possibility is 
in the middle: reasonably well-separated but partially overlapping swamplands that allow for independent tests of different 
fundamental theories of gravity against the existing wealth of experimental data, e.g., on particle physics, 
black hole physics and cosmology.  A subcase of this scenario is when the string landscape encompasses all other  landscapes; this is known as the string lamppost principle \cite{Montero:2020icj}. It implies that the string swampland is contained in all other relative swamps and hence coincides with the absolute swampland.
 
Against this background, the
purpose of the present note is to review the status of three prominent swampland conjectures 
in string theory and asymptotic safety: the \textit{no-global-symmetries}, \textit{weak gravity}, and \textit{de-Sitter} 
conjectures. Along the way, we will entertain mostly what we depicted as scenario II in Fig.~\ref{fig:sketch}, in view of
understanding the universality of swampland criteria. We outline pertinent research directions in the outlook.

\section{Background concepts}
\label{sec:Concepts} 

While we will compare landscapes/swamplands of EFTs based on their shared universal principles, we wish to briefly 
recall that they play a rather different role in string theory and in  asymptotically safe gravity.

\subsection{The landscape and swampland in asymptotic safety}
\label{sec:AS-Swampland}

Asymptotically safe gravity is based on an interacting ultraviolet (UV) fixed point of the Renormalisation Group 
(RG) flow in quantum gravity. By now there is compelling evidence for the existence of the fixed point in a purely 
gravitational theory and in some gravity-matter theories, see \cite{Eichhorn:2022gku, Pawlowski:2023gym} for recent 
reviews. A four-dimensional effective field theory is defined by: (i) its matter content, (ii) its interaction terms, 
(iii) the infrared (IR) values of its coupling constants. Such a theory is in the relative 
landscape of asymptotically safe 
gravity iff its RG flow emanates from an asymptotically safe UV fixed point, see \cite{Basile:2021krr, Platania2024} for examples of a landscape constructed in this way. If this fixed point has $n$ relevant 
perturbations, we are left with an $n$-dimensional landscape in theory space. In addition, the landscape can contain 
EFTs that result from distinct UV fixed points (some of which may correspond to different vacua). For instance, there 
is evidence that there is both a UV fixed point with only gravitational degrees of freedom, as well as a fixed point 
with gravitational and matter degrees of freedom \cite{Dona:2013qba, Meibohm:2015twa}. In addition, there may be (some) freedom in 
the choice of spacetime dimensionality, also in the UV.

\subsection{The landscape and swampland in string theory}
\label{sec:String-Swampland}

In contrast, the 4-dimensional EFTs that make up the string landscape \cite{Susskind:2003kw} are more usefully thought 
of as arising from a single, unique UV theory that is placed in different backgrounds.\footnote{More precisely, there 
is a handful of ways of defining UV-finite microscopic interactions, but these end up being related by duality upon 
compactification and symmetry breaking.} Each background is characterised by its own (Kaluza-Klein) scale, above which 
the universal higher-dimensional theory begins to take over, and a set of fields and interactions that run 
down to ``very low'' energies, unless protected by supersymmetry. The UV theory itself is not a field theory.
The AdS/CFT correspondence can be construed \cite{Banks:2003vp, Banks:2012hx} to imply that different asymptotic 
behaviour, dictated in particular by the cosmological constant, in fact corresponds to different theories, and not 
to different vacua within a single theory. This presents an obstacle to certain cosmological considerations, but 
not to the discussion of the resulting EFTs as a set, which famously is enormously large. The relative swampland 
of string theory are then those EFTs that cannot be UV completed in this fashion.

\section{No-global-symmetries conjecture}
\label{sec:No-Global_symm}

\subsection{Conjecture} 
\label{sec:ConjectureI}

The 
conjecture
is that quantum gravity forbids global symmetries \cite{Banks:1988yz, Giddings:1987cg, Lee:1988ge, Abbott:1989jw, Coleman:1989zu, Kamionkowski:1992mf, Holman:1992us, Kallosh:1995hi, Banks:2010zn}. 
Instead, the only consistent symmetries are gauged. This should hold both 
for continuous and discrete symmetries,
cf.~e.g.~\cite{Krauss:1988zc} for the distinction between global and gauged discrete symmetries.\\
The conjecture is closely tied to black-hole physics: Given a matter field that transforms under a symmetry group with infinitely many irreducible representations, the collapse of an appropriate initial configuration of this field can form a black hole that carries an arbitrarily complicated representation.  
One may consider three assumptions concerning black holes in general and the special black holes above in particular:
\begin{enumerate}[leftmargin=*,topsep=1ex,itemsep=.5ex,partopsep=.5ex,parsep=.5ex]

\item[(i)] Black holes evaporate according to Hawking's description. 
\item[(ii)] Black holes do not leave behind remnants that have an entropy much larger than the Bekenstein-Hawking-entropy.
\item[(iii)] The path integral contains contributions from virtual black-hole configurations.

\end{enumerate}
The first two assumptions are sufficient to conclude that continuous global symmetries are violated. 
This violation is 
actually
not a conjecture but rather a direct consequence of the low energy effective theory, which we assume correctly describes 
black hole evaporation.\footnote{
See 
however \cite{Dvali:2020wft}.
}
If we include the third assumption\footnote{
In supersymmetric field and string theory, microscopic BPS states of extremal black holes famously do contribute to the evaluation of physical observables \cite{Strominger:1995cz,Gopakumar:1998jq}.}
one can say more:
The effective action for matter that arises by performing the semi-classical path integral for gravity contains interactions that violate the global symmetry. Local symmetries are allowed because black holes carry `hair' described by the appropriate generalisation of the  
Kerr-Newman
solution. Discrete symmetries which only have finitely many irreducible representations, such as $\mathbb{Z}_2$ (which only has one nontrivial representation) are not excluded by the black-hole argument. However, many of the early arguments in favour of the conjecture involve euclidean wormholes. If these are included in the path integral, they carry away charge and violate both continuous and discrete global symmetries, see \cite{Hebecker:2018ofv} for a review.

Based on the above, an obvious question with relevance to phenomenology is: If exact global symmetries are forbidden, what is the maximal quality an approximate global symmetry can have?
For the specific case of axionic shift symmetries, this has been intensely debated in the context of large-field inflation, on the basis of the weak-gravity conjecture for axions (see e.g.~\cite{delaFuente:2014aca, Rudelius:2015xta, Montero:2015ofa, Brown:2015iha, Bachlechner:2015qja, Hebecker:2015rya}). In the non-inflationary axionic context, the reader may consult the review \cite{Reece:2023czb}. For linearly realised, continuous symmetries, arguments using the weak-gravity conjecture, gravitational K3 instantons, and assumptions about the violation rate by black holes in a thermal plasma have been given \cite{Hebecker:2019vyf, Fichet:2019ugl, Daus:2020vtf}. Moreover, for the important class of approximate global symmetries following from spontaneously broken gauge symmetries, a derivation based on the weak-gravity conjecture has been provided \cite{Daus:2020vtf}. Consistently with early wormhole arguments \cite{Giddings:1987cg, Lee:1988ge, Abbott:1989jw, Coleman:1989zu} and with the recent wormhole-based calculation in AdS \cite{Bah:2022uyz}, all this suggests that the minimal violation is quantified by $\exp(-M_P^2/\Lambda^2)$, with $\Lambda$ the cutoff of the low-energy effective theory \cite{Fichet:2019ugl, Daus:2020vtf}. To be more precise, the claim is that the matter-field effective action should contain symmetry-violating terms with Wilson coefficients $\exp(-M_P^2/\Lambda^2) /\Lambda^{d-4}$, where $d$ is the dimensionality of the corresponding operator.

\subsection{Status in string theory}

Maybe the first clear statement of the conjecture appears in the string theoretic context \cite{Banks:1988yz}. It comes together with a derivation in string perturbation theory, based on worldsheet logic: Any global symmetry of the low energy effective theory is also a global symmetry on the worldsheet. The corresponding worldsheet Noether current can be shown to have all the features making it a suitable vertex operator creating a massless vector. This vector is the gauge boson of the symmetry in question, which is hence not global \cite{Banks:1988yz}.

A key limitation of the above argument, apart from its perturbative nature, is the restriction to continuous symmetries. This has been recently remedied \cite{Harlow:2018tng} using AdS/CFT \cite{Maldacena:1997re, Aharony:1999ti}. Given that AdS/CFT, while not intrinsically stringy, has its most solid roots in string theory, we discuss this in the present section: The basic idea in  \cite{Harlow:2018tng} is to consider quantum gravity in AdS and to postulate that some local operator transforms non-trivially under a global symmetry. This transformation can then also be realised by an operator in the boundary CFT. Moreover, that action may be split into actions on separate regions (a plausible feature of CFTs called `splittability').
These regions can be chosen small such that the part of the AdS bulk affected by each region (its entanglement wedge) is also small. Then, in total, there remains a part of the bulk where the local operator we started with cannot be affected, in contradiction to our assumed global symmetry. A remaining limitation is that this argument only works in AdS.

\subsection{Status in asymptotic safety}
\label{sec:GlobalSyms-AS}

Gravitational fluctuations generate interactions for matter fields in minimally (no matter self-interactions) coupled matter-gravity systems; their couplings accordingly have nonzero fixed-point values. In the absence of anomalies, these interactions carry the same global symmetries as the minimally coupled theory. This is a \textit{mathematical} consequence of the functional flow for the effective action, which has been corroborated within numerous explicit calculations, see \cite{Eichhorn:2022gku} for references on scalar, fermionic and vector field theories with various global symmetries. The above is a necessary, but not sufficient condition that an asymptotically safe fixed point respects global symmetries, as the vacuum (the solution of the quantum equation of motion) may break the global symmetry or the full theory may not even have a fixed point. 

The most extensively explored case is that of scalar fields, and the simplest example is that of a single real scalar field. 
The minimally coupled version only features a kinetic term with $\mathbb{Z}_2$ symmetry 
as well as shift-symmetry. Metric fluctuations only generate self-interactions and non-minimal scalar-curvature couplings with
$\mathbb{Z}_2$- and shift-symmetry \cite{deBrito:2021pyi, Laporte:2021kyp}, see \cite{Eichhorn:2022gku} for further references. Generalising to $O(N)$ symmetry, metric fluctuations only generate $O(N)$ preserving interactions \cite{Labus:2015ska, deBrito:2021pyi}. 

By contrast, for discrete $\mathbb{Z}_{n}$ symmetries with $n>4$, interactions stay irrelevant at the free fixed point and no interacting fixed point was found \cite{Ali:2020znq}.
The corresponding low-energy theory can therefore not have this discrete symmetry, but instead is a trivial, i.e., non-interacting theory in the scalar sector. Thus, the asymptotically safe UV completion can indeed clash with a discrete global symmetry; however, the clash is not resolved by gauging the symmetry but by trivialising the theory (at least in the matter sector).

For theories with fermions, global chiral symmetries are well explored and metric fluctuations only generate interactions that preserve these symmetries, as long as no explicit breaking is introduced via the regulator term, see, e.g., \cite{Eichhorn:2011pc, Meibohm:2016mkp, Eichhorn:2016vvy, Daas:2021abx,deBrito:2023kow}.
There is also support for the existence of global symmetries from lattice simulations. In Euclidean Dynamical Triangulations, fluctuations of spacetime preserve an auxiliary U(1) symmetry for fermions which evolves into the chiral symmetry in the continuum limit \cite{Catterall:2018dns}.

There are two possible explanations for these results: Taken at face value they would imply that the no-global-symmetries conjecture does not hold in asymptotic safety, and some of the three assumptions listed above would have to be violated. Indeed, there are hints that black-hole evaporation may lead to remnants \cite{Bonanno:2006eu, Pawlowski:2018swz, Platania:2023srt}, explicitly invalidating assumption (ii). 
In addition, a suppression mechanism for (singular) black-hole configurations in the path integral exists and is tied to higher-order curvature invariants in the dynamics  \cite{Borissova:2020knn}. Such a mechanism implies that, at least for some classes of black holes, assumption (iii) may also not hold in asymptotic safety.

Alternatively, the fixed points with global symmetries may
in principle
be artefacts of approximations in the calculations. Most importantly, given the Euclidean setup of most calculations to date, black-hole configurations, or more generally configurations with 
apparent/event horizons, are likely not accounted for properly. A similar statement might apply to gravitational instantons, see e.g.~\cite{Hamada:2020mug}. Finally, calculations to date may not fully account for anomalies.

\subsection{Future perspectives}

To achieve unitary black-hole evaporation, some degree of non-locality may be necessary, as evidenced by recent progress on obtaining the Page curve for Hawking radiation in the context of AdS/CFT \cite{Almheiri:2020cfm}. Invoking nonlocality may, however, also provide a mechanism to make information on global charges available outside the horizon. Understanding the interplay of unitarity, locality and global symmetries in black-hole evaporation is therefore crucial, see \cite{Harlow:2020bee} for arguments that connect unitarity to non-existence of global symmetries.

Going forward, there are several distinct ways to further test the no-global-symmetries conjecture in asymptotic safety. 
First, in functional Renormalisation Group studies, working in Lorentzian signature may be key to properly account for all 
configurations in the path integral, see \cite{Fehre:2021eob} for recent progress in that direction.
Second, evidence could be strengthened using non-perturbative numerical simulations, e.g., in the causal (or Euclidean) 
dynamical triangulations framework, see \cite{Ambjorn:2024pyv}. We caution that spacetime topology is restricted in these 
simulations, in order to obtain a phase with an extended four-dimensional universe. We also speculate that asymptotic 
safety may be possible in settings with as well as without spacetime topology change. These may differ regarding the 
status of global symmetries.
 
Broadening our view, we suggest to explore the fate of global symmetries in other approaches to quantum gravity, including, 
e.g., Loop Quantum Gravity, where the coupling of matter fields has been considered, e.g., in \cite{Mansuroglu:2020dga}.

\section{Weak-gravity conjecture}
\label{sec:WeakGravity}

\subsection{Conjecture}
\label{sec:WG-Conjecture}

On the one hand, we have argued that global symmetries are forbidden. On the other hand, one may think of them as emerging from gauge symmetries in the limit of vanishing coupling. Hence, guided by the somewhat vague but appealing principle that `physics is 
continuous', we expect that quantum gravity prescribes a smallest consistent gauge coupling. Based among others on many string-theoretic examples, such a `quantitative version of the no-global symmetry conjecture' has indeed been proposed. It is known as the weak-gravity conjecture \cite{ArkaniHamed:2006dz} and states roughly that 
\begin{align}
\frac{\Lambda}{M_{\rm Planck}} \lesssim
g\,,
\end{align}
where $\Lambda$ is the ultraviolet cutoff of the low-energy effective theory. Put differently, the limit of vanishing $U(1)$ gauge coupling is sick, as in such a limit the validity regime of the low-energy effective theory goes to zero.

This can be cast in a more precise form, which is related but not strictly equivalent to the above: there should always exist a charged state in the theory the mass of which in Planck units is smaller than its charge, 
\begin{align}
\frac{m}{M_{\rm Planck}} \leq e\,q\,\sqrt{2}\,.\label{eq:WGC}
\end{align}
Here $e$ is the gauge coupling and $q$ is the integer charge as it is conventionally measured by the surface integral of the dual field strength, cf.~e.g.~the recent review \cite{Harlow:2022ich}. More simply put, gravity should be weaker than the electromagnetic force. This last reinterpretation becomes obvious by noting that \eqref{eq:WGC} is equivalent to the statement that $eq/m$ is larger than the charge to mass ratio of an extremal black hole and that for extremal black holes gravitational attraction precisely compensates electromagnetic repulsion.

Generalisations to $p$-form gauge theories, including axions and implications for inflation, have been discussed (see e.g.~\cite{delaFuente:2014aca, Rudelius:2015xta, Montero:2015ofa, Brown:2015iha, Bachlechner:2015qja, Hebecker:2015rya}). It is conceivable that a model satisfying the weak-gravity conjecture at high scales violates it after some form of Higgsing in the infrared \cite{Hebecker:2015rya, Saraswat:2016eaz}.

\subsection{Status in string theory}
\label{sec:StatusWGString}

As already mentioned, a key motivation for the weak-gravity conjecture is the observed correlation between small 
gauge coupling and low cutoff scale (or corresponding light states) in explicit string constructions. Stringy examples, 
in particular the heterotic theory, have played a central role in the motivation \cite{ArkaniHamed:2006dz} and initial 
study of the conjecture. A key argument for the conjecture uses the same reasoning as outlined in the no-global-symmetries 
context above: The fundamental object is the global symmetry current on the worldsheet that provides the vertex operator 
for the space-time gauge boson. This current living in the ``compact part'' of the CFT, it integrates to a compact worldsheet 
boson, and charged vertex operators lead to space-time states required by the weak-gravity conjecture. In particular, they 
have a higher-dimensional stringy origin. The existence of such
vertex operators had been sketched in\cite{ArkaniHamed:2006dz}. It was recently proven rigorously in  
\cite{Heidenreich:2024dmr}, based on a careful analysis of modular invariance and together with a calculation of the
repulsive long-range forces, albeit only for the perturbative bosonic string. Ramond-Ramond charges being carried by D-branes, the superstring will require a somewhat different set of tools. For holographic arguments for the weak gravity conjecture see e.g.~\cite{Harlow:2015lma, Montero:2016tif, Montero:2018fns}.

Both for the proof as well as, more generally, for the conceptual understanding of the 
role of the weak-gravity conjecture in the swampland program, it is crucial that the key claims are not merely about 
a single charged state being light. On the contrary, string-theoretic as well as Kaluza-Klein logic imply the presence 
of a whole tower or lattice of charged states 
\cite{Heidenreich:2015nta, Heidenreich:2016aqi, Andriolo:2018lvp, Grimm:2018ohb, Lee:2018urn, Corvilain:2018lgw, Gendler:2020dfp}. Important roles are 
furthermore played by the BPS property enjoyed by the lightest charged states in many cases \cite{Ooguri:2016pdq} 
and by the no-force condition between charged states or extremal  black holes \cite{Palti:2017elp, Lee:2018spm, Heidenreich:2019zkl}. 
We also note that stringy counterexamples have been essential in excluding naively appealing but in hindsight too 
strong forms of the conjecture \cite{Heidenreich:2016aqi}. For example, it is not true that the lightest charged 
particle or the particle with the smallest non-zero charge always satisfy \eqref{eq:WGC}.

A relatively weak but very interesting and accessible form of the conjecture arises by applying the inequality \eqref{eq:WGC} to black holes. One finds the requirement $(M/Q)\geq (M/Q)_{\rm extremal}$, where $M$ and $Q$ are ADM mass and charge and $(M/Q)_{\rm extremal}$ is the corresponding ratio for extremal black holes in the large-mass limit. The inequality may then be enforced by the effect of higher-derivative terms, which can be explicitly analysed in string theory as well as in classes of effective field theories \cite{Kats:2006xp, Cheung:2018cwt, Hamada:2018dde, Arkani-Hamed:2021ajd}.

\subsection{Status in asymptotic safety}
\label{sec:StatusWGAS}

To interpret the weak-gravity conjecture in asymptotic safety, we have to decide at which scales it should hold. To begin with, the inequality \eqref{eq:WGC} must at least hold at IR values of the RG scale. A stronger condition is to require \eqref{eq:WGC} to hold true at all RG scales, see \cite{deAlwis:2019aud}. This stronger requirement also constrains fixed-point values. From \eqref{eq:WGC}, it follows that the Abelian gauge coupling must not vanish at the UV fixed point. To order $e^3$, the beta-function for the Abelian gauge coupling is
\begin{align}
\beta_e= -f_g\, e + \beta_0\, e^3,
\end{align}
where $\beta_0$ is the one-loop coefficient and $f_g>0$ is a function of the gravitational couplings encoding gravitational fluctuations, \cite{Harst:2011zx,Eichhorn:2017lry}. This beta function admits two different fixed points: $e_{\ast}=0$, at which the Abelian gauge coupling is a relevant perturbation of the fixed point and $e_{\ast}>0$, at which it is an irrelevant perturbation, connected to a unique IR value. If we demand
that the weak-gravity conjecture holds at all scales, it excludes the first, non-interacting fixed point, $e_{\ast}=0$. 
The conjecture might hold at the second, interacting fixed point $e_{\ast}>0$; however 
the resulting prediction for the coupling might not be in agreement with experimental results on the coupling strength of electromagnetism \cite{Eichhorn:2017lry}. The systematic theoretical uncertainties associated to the calculation of the IR value of $e$ are currently too large to decide this point. Regarding the mass of the charged state, it is important to distinguish fermions from scalars.
At an interacting fixed point for the gauge coupling, shift-symmetry for a charged scalar field is broken and thus the mass $m_s$ of a scalar field also has an interacting fixed point $m_{s\,\ast}\neq 0$, see, e.g., \cite{Eichhorn:2019dhg}. Both fixed-point values, $e_{\ast}$ and $m_{s\,\ast}$, depend on the gravitational fixed-point values. The inequality \eqref{eq:WGC} thus constrains the gravitational fixed-point values \cite{deAlwis:2019aud}. In contrast, for fermions, chiral symmetry continues to protect the mass, such that $m_{f\,\ast}=0$ holds, given the present status of results regarding global symmetries, see previous section.

If one requires the weak-gravity conjecture only in the deep IR, then one may start from the asymptotically free fixed point $e_{\ast}=0$ and use the associated relevant direction to select RG trajectories which satisfy the conjecture in the IR; both scalar and fermion masses are relevant parameters, as long as asymptotically safe gravity is not too strongly coupled.

We note that the above pertains to the existence of a single light charged state, not a tower of light charged states, as a difference to the discussion in string theory.

 A different possibility to approach the weak-gravity conjecture in asymptotic safety arises \cite{Basile:2021krr,Platania2024} because, as explained above, higher derivative terms affect the mass-to-charge ratio of small black holes \cite{Kats:2006xp, Cheung:2018cwt, Hamada:2018dde, Arkani-Hamed:2021ajd}.  Such terms, in particular higher-curvature corrections, are generically present in asymptotic safety. 
 In \cite{Platania2024}, they are found to lead to Planck-scale suppressed violations of the conjecture for RG flows that start from a fully interacting fixed point. A previous study \cite{Basile:2021krr} that accounted for fewer terms in the effective action and started from a fixed point that is asymptotically free in the curvature-squared couplings found that the conjecture was satisfied. This suggests the possibility that different asymptotically safe UV completions may have different swamplands.
 \\

\subsection{Future perspectives}

Lifting of some of the approximations going into the current results in asymptotically safe gravity, most importantly, truncations of the dynamics, will provide a clearer picture of whether the weak-gravity conjecture holds in the UV and on which subset of trajectories it holds in the IR. In addition, there is currently no indication that the cutoff of the EFT moves down and a tower of states must exist if the gauge coupling becomes small. This may, however, apply to asymptotic safety in a setting with extra dimensions, which remains to be investigated.

There are also attempts to prove the weak-gravity conjecture in the context of 
positivity bounds \cite{Cheung:2018cwt,Hamada:2018dde}, which constrain 
effective field theories based on requiring unitarity, (micro)causality, 
locality (in a form compatible with the nonlocality present in string theory) 
and Lorentz invariance of the UV completion. At the time of writing, these 
conditions are insufficient to derive the weak-gravity conjecture 
\cite{Henriksson:2022oeu}, making explicit tests of the conjecture in 
quantum-gravity approaches other than string theory even more pressing.

\section{De-Sitter conjecture} 
\subsection{Conjecture} 

Even in semiclassical gravity, without a UV completion, many authors have questioned the possibility of eternal de Sitter space and attempted to bound its lifetime. Arguments have been given involving the backreaction of quantum fluctuations, quantum decoherence and others. A very incomplete list of references is \cite{Ford:1984hs, Mottola:1984ar, Antoniadis:1985pj, Tsamis:1992sx, Mukhanov:1996ak, 
Maldacena:2000mw, Goheer:2002vf, Polyakov:2007mm, 
Dvali:2013eja, Dvali:2017eba, Matsui:2018iez}.

\subsection{Status in string theory}
Based on the 10d superstring, many well-controlled AdS or Minkowski vacua can be constructed. Since one expects the string landscape to be connected, stringy de Sitter vacua can presumably always decay and are hence at best metastable. The technical challenge of creating even such metastable constructions has lead to the stronger proposal that `string theory has no de Sitter vacua' \cite{Danielsson:2018ztv}. A more quantitative formulation with far-reaching cosmological implications takes the form of the so-called $V'/V$ conjecture \cite{Obied:2018sgi}, excluding locally flat potentials. While such a strong conjecture appears to be in conflict with experiment \cite{Denef:2018etk}, a weaker formulation employing a combined 
constraint on $V'/V$ and $V''/V$ may hold \cite{Ooguri:2018wrx,Garg:2018reu}.

In the asymptotics of field space, derivations of the weaker conjecture above have been suggested \cite{Ooguri:2018wrx, Hebecker:2018vxz}, but these involve non-trivial assumptions. As a more general justification for conjectures against de Sitter, one may recall that string theory is fundamentally defined on the basis of the S-matrix, which is not available in de Sitter space (see e.g.~\cite{Freivogel:2004rd,   Dvali:2013eja, 
Dine:2020vmr, Berezhiani:2021zst, Banks:2023uit} and refs.~therein). However, it is not clear that this can rule out de Sitter as a metastable state. It appears that a general derivation of the de Sitter conjecture in string theory is not in sight. Thus, the question about the parametric control of the few explicit de Sitter constructions, in particular \cite{Kachru:2003aw, Balasubramanian:2005zx}, is more pressing than ever. This is a longstanding problem, with some of the most critical recent issues being related to the non-trivial interplay of the Calabi-Yau volume, the depth of the warped throat with its anti-D3-brane uplift, and curvature corrections making the latter potentially unstable \cite{Carta:2019rhx, Gao:2020xqh, Junghans:2022exo, Gao:2022fdi, Hebecker:2022zme}. On the positive side, we  
mention recent progress in understanding the complex-structure-based $F$-term uplift \cite{Krippendorf:2023idy} as an alternative to anti-D3-branes. 

Phenomenologically, quintessence type models \cite{Wetterich:1987fm, Ratra:1987rm, Caldwell:1997ii} allow for a description of today's cosmic acceleration even if the conjecture holds. However, finding a {\it realistic} quintessence model in string theory does not appear to be simpler compared to de Sitter, see e.g.~\cite{Cicoli:2012tz, Cicoli:2018kdo, Hebecker:2019csg}.

\subsection{Status in asymptotic safety}

The gravitational background solution of the quantum equations of motion in asymptotically safe theories carries an inherent scale dependence. For example, it may be AdS-like for asymptotically small distances, while it is de Sitter-like for large (cosmological) distances. In the RG approach to asymptotically safe theories this is encoded in a rather simple form: there, we distinguish the UV value of the cosmological constant in the UV fixed-point regime 
from its IR value. The two are connected by an RG trajectory, which emulates the length-scale dependence of the full gravitational background. Here the cosmological constant ``runs" as a function of an IR cutoff scale in the path integral, which is in general different from the ``running" with a physical (momentum) scale \cite{Bonanno:2020bil}. 
Only its IR value is constrained by observations in cosmology. The UV value, i.e., the fixed-point value, can be positive or negative, depending on the matter content of the theory, see e.g.~\cite{Dona:2013qba, Meibohm:2015twa}, for recent reviews see \cite{Eichhorn:2022gku, Pawlowski:2023gym}. The cosmological-constant term is a relevant perturbation of the fixed point and a given IR cosmological constant lies on a specific trajectory that emanates from the UV fixed point. Crucially, positive IR values can be reached, irrespective of whether the fixed-point value itself is positive or negative, see, e.g., \cite{deAlwis:2019aud} for an explicit example. Within an analytic continuation of the Euclidean results to Lorentzian signature, a positive value of the cosmological constant appears to be compatible with asymptotic safety. Such an analytic continuation from Euclidean space poses a significant challenge in QFTs, leave alone quantum gravity. Typically one resorts to (ill-conditioned) extrapolations, for first direct computational results see \cite{Fehre:2021eob}. A positive value of the cosmological constant is, however, not sufficient to show the compatibility of asymptotic safety with de Sitter solutions. Instead, one must show that de Sitter remains a solution to the effective equations of motion in the presence of higher-curvature terms and at large curvatures; see \cite{Dietz:2012ic, Falls:2016wsa, Christiansen:2017bsy, Falls:2018ylp, Mitchell:2021qjr} for explicit studies accounting for $f(R)$ terms. 

Closer in spirit to string theory is the setting with a real scalar degree of freedom with $\mathbb{Z}_2$-symmetry and dynamical dark energy. 
In this setting, one may question whether asymptotic safety even singles out or prefers flat potentials, or whether they maybe only be achieved with fine-tuning. 

We first consider this question within an expansion around vanishing scalar fields or small curvatures. Then, the scalar effective potential reads $V[\phi] = \sum_{i=1}^{\infty}\lambda_i \phi^{2i}$, with the mass parameter $\lambda_1$, the quartic coupling $\lambda_2$ and higher-order couplings. It turns out that the quantum-gravity correction to their RG scaling screens the interactions and thus lowers the scaling exponents of the couplings at the UV fixed point with $\lambda_{i\, \ast}=0$. In this scenario, the quartic coupling is an irrelevant perturbation of the fixed point, e.g., \cite{Narain:2009fy, Eichhorn:2020sbo}. The mass is always less relevant than its canonical scaling suggests and can even become irrelevant, if gravitational fluctuations are strong enough, see e.g.~\cite{Narain:2009fy,Wetterich:2016uxm,Eichhorn:2020sbo}. This exciting scenario is corroborated in studies that take into account the full potential including the large field behavior, \cite{
Henz:2016aoh, Pastor-Gutierrez:2022nki, deBrito:2023myf, Wetterich:2019rsn}. These works even suggest that asymptotic safety enforces globally flat scaling solutions with zero to two relevant parameters.\\ 
Finally, in \cite{Basile:2021krr}, the potential for the scalar degree of freedom encoded in the $R^2$ term was explored; finding that the refined de Sitter conjecture holds only in part of the asymptotically safe landscape; in line with the existence of relevant directions that imply free parameters in the IR scalar potential.

\subsection{Future perspectives}
Asymptotically safe theories may serve as (low energy) effective theories below some trans-Planckian physics scale in the scaling regime around their respective UV fixed points. Above this scale they may lack the full physics of the underlying fundamental quantum gravity theory,\footnote{One 
may also question the physical interpretability of this regime altogether \cite{Dvali:2010bf}.
}
while still being a valid QFT. In any case, this set-up includes asymptotically safe theories whose gravitational background  solution of the equations of motion is AdS-like at small trans-Planckian distances, while it is de Sitter-like at large scales. This scenario includes QFTs which are low-energy EFTs of a string theory, as proposed in \cite{deAlwis:2019aud}, and opens the door to string theory set-ups with a trans-Planckian string scale.  
A first step towards the exploration of these exciting scenarios, would be to consider string theory on a scale-dependent background. Potentially, this may even reconcile string theory with de Sitter backgrounds. 

Given cosmological surveys of baryon acoustic oscillations and the resulting constraints on the equation of state of dark energy, observational statements on the flatness of scalar potentials in cosmology appear to be in sight \cite{Heisenberg:2018yae}. Thus, it is essential to understand whether de Sitter and near de Sitter spacetimes lie in an absolute swampland or in the relative swampland of a subset of quantum gravity theories, such that large-scale cosmological observations can constrain the small-scale structure of spacetime.

\section{Conclusions and outlook}
Given the current results, the universality of some swampland conjectures appears questionable. 
This is clearly a promising perspective for phenomenological tests of quantum gravity and calls for a more in-depth study of the above as well as additional conjectures.
Here, we sketch some promising future directions for research.

From the string theory perspective, a very important conjecture is the so-called Distance Conjecture \cite{Ooguri:2006in}. In its simplest form, it says that when moving to infinity in field space the quantum gravity cutoff comes down exponentially. In $d<10$, this field-space infinity is always associated with decompactification, such that the decreasing cutoff is related to the KK scale \cite{Lee:2019wij}. 
A similar situation would arise in asymptotic safety if one were to define the theory in $d>4$ and compactify, see, e.g., \cite{Litim:2003vp,Ohta:2013uca, Dona:2013qba, Eichhorn:2019yzm}  for first studies of asymptotic safety in $d>4$.
However, there appears to be the alternative possibility of simply adding a shift-symmetric scalar to the spectrum. 
As far as we know, this does not in general endanger the UV fixed point -- given the status of global symmetries reviewed above -- but it introduces a field-space infinity without decreasing cutoff. 
Here, FRG studies that do not rely on small-field expansions, and fully account for induced self-interactions and non-minimal interactions may be important to determine whether a shift-symmetric scalar potential is viable; our comments on how to further test the no-global-symmetries conjecture also apply.

In asymptotically safe gravity, work has to date focused on investigating whether or not specific particle-physics models lie in the asymptotically safe landscape or in the theory's swampland, with promising results regarding the Standard Model \cite{Eichhorn:2017ylw, Alkofer:2020vtb, Pastor-Gutierrez:2022nki} and beyond, see \cite{Eichhorn:2022gku} for a review.  
Generalizing these results to the formulation of new, asymptotic-safety inspired swampland conjectures has so far not been attempted.

Broadening our perspective, the study of swampland conjectures in two (or more) candidate quantum gravity theories also raises the question of a deeper relation between these theories. 
Although, at a first glance, asymptotically safe gravity and string theory are based on different assumptions, one should carefully distinguish between the entities used in formulating the theory and the resulting physical observations. As proposed in \cite{deAlwis:2019aud}, the two theories may be compatible if asymptotic safety is realized as an intermediate regime of a string theory. 
In such as setting, asymptotic safety can in principle also serve as a mechanism that generates universality. 
This mechanism relies on the infrared-attractive directions of an asymptotically safe fixed point, along which trajectories can be attracted towards an intermediate scaling regime. 
Depending on the compatibility of asymptotic safety with supersymmetry \cite{Dona:2014pla, Montero:2024sln}, such a regime may be achieved above or below the supersymmetry breaking scale.

Given the central role of the holographic principle in the string-theoretic understanding of gravity, what is its role 
in asymptotic safety? We highlight two aspects that deserve further research.
First, one could develop the asymptotically safe perspective on the AdS/CFT correspondence. In its most concrete realization,
this is a duality between maximally supersymmetric Yang-Mills theory and type IIB string theory in AdS space. The basic
dictionary however depends only on the semi-classical approximation and is essentially independent of the UV completion.
It would be interesting to understand whether asymptotically safe gravity can be defined on an asymptotically AdS 
spacetime, and possibly differs in its quantum predictions for the dual field theory. This should in particular 
significantly sharpen the basic conundrum posed by the holographic principle regarding the number of degrees of freedom.
Asymptotic safety is by definition scale-symmetric and might even exhibit full conformal symmetry. However, a 5d
CFT in the bulk has many more degrees of freedom than a 4d CFT on its boundary \cite{Shomer:2007vq}. Already, 
there are first indications that this counting may be too naive in a quantum gravitational setting, in which dynamical 
dimensional reduction to lower spacetime dimensionality occurs in the fixed-point regime, cf.~e.g.~the discussion in \cite{Bonanno:2020bil}.\footnote{See also the non-standard correspondence between 5d AdS and a 4d asymptotically safe theory including gravity proposed in \cite{Ferrero:2022dpk}.}

Conversely, one should wonder what the holographic principle really says away from the semi-classical regime, and 
independently of the assumed UV completion. Indeed, just like most of the swampland conjectures that we have discussed in 
this paper, the most stringent arguments for the holographic principle rely on the semi-classical properties of black
holes and their realization in string theory. However, the structure and role of black holes at the Planck scale might 
differ significantly in asymptotically safe gravity, e.g., due to higher-order terms in the dynamics.

To close, we observe that a completely universal swampland is unrealistic to expect. Instead, it is likely that there is an absolute swampland, but also non-empty intersections between the relative swampland of one quantum-gravity theory and the landscape of another theory. 
We thus conjecture that some
parts of the relative swamplands are directly linked to distinct physical UV properties of different theories and therefore differ. Delineating the absolute swampland as well as its relation to the relative swamplands is thus key to testing quantum-gravity theories through experiments and observations.

{\bf Acknowledgements} 
We are grateful to M.~Montero and A.~Platania for discussions and comments on our manuscript.
This work is funded by the Deutsche Forschungsgemeinschaft (DFG, German Research Foundation) by Germany’s Excellence Strategy EXC 2181/1 - 390900948 (the Heidelberg STRUCTURES Excellence Cluster). AE is supported by a research grant (29405) from VILLUM Fonden.
JW thanks the International Centre for Mathematical Sciences, Edinburgh, for support and hospitality during the ICMS 
Visiting Fellows programme where this work was completed. This work was supported by EPSRC grant EP/V521905/1.

\bibliographystyle{myutphys}
\bibliography{refs}

\end{document}